\documentclass[aps,prc,tightenlines,showpacs,preprint,showkeys,%
amssymb,byrevtex,nofootinbib,superscriptaddress]{revtex4}  

\usepackage{epsfig,bm,amsmath}

\begin{document}



\title{Spin asymmetries in $\bm{\gamma N \to \overline{K}^* \Theta^+}$}


\author{Yongseok Oh}%
\email{yoh@physast.uga.edu}

\altaffiliation[Present address: ]{Department of Physics and Astronomy,
University of Georgia, Athens, GA 30602, U.S.A.}

\affiliation{Thomas Jefferson National Accelerator Facility, 12000
Jefferson Avenue, Newport News, VA 23606, U.S.A.}

\affiliation{Institute of Physics and Applied Physics,
Yonsei University, Seoul 120-749, Korea}

\author{Hungchong Kim}%
\email{hung@phya.yonsei.ac.kr}

\affiliation{Institute of Physics and Applied Physics,
Yonsei University, Seoul 120-749, Korea}

\author{Su Houng Lee}%
\email{suhoung@phya.yonsei.ac.kr}

\affiliation{Institute of Physics and Applied Physics,
Yonsei University, Seoul 120-749, Korea}

\date{\today}


\begin{abstract}

The photoproduction processes of the exotic $\Theta^+(1540)$ baryon and
the $\overline{K}^*$ meson from the nucleon targets, i.e.,
$\gamma n \to K^{*-} \Theta^+$ and $\gamma p \to \bar{K}^{*0} \Theta^+$,
are investigated in a hadronic model.
We consider $K$ and $K^*$ exchanges as well as the $s$ and $u$ channel
nucleon and $\Theta$ terms.
Various spin asymmetries together with cross sections are first computed in
order to study the production mechanisms and the parity of the
$\Theta^+(1540)$ baryon.
Within the uncertainties arising from the model-dependence of the production
mechanisms and several coupling constants, we find that some target-recoil
double spin asymmetries, $C^{\rm TR}_{xx'}$ and $C^{\rm TR}_{xz'}$, are
sensitive to the parity of $\Theta^+$.
In addition, the parity asymmetry of this reaction on the neutron target,
which can be obtained by analyzing $K^*$ decay distribution, is found to be
useful to estimate the $K^* N \Theta$ coupling.

\end{abstract}


\maketitle

\section{Introduction}

Since the discovery of the $\Theta^+(1540)$ by the LEPS Collaboration at
SPring-8 \cite{LEPS03}, there have been a lot of experimental and
theoretical studies for the exotic baryons.
Experimentally, the observation of the $\Theta^+(1540)$ has been
confirmed by many other experiments 
\cite{DIANA03,CLAS03-b,SAPHIR03,CLAS03-c,ADK03,CLAS03-d,HERMES03},
of which results are summarized in Table~\ref{tab:expt}.
The decay width of the $\Theta^+(1540)$ is found to be small but,
due to the limitation in the detector resolution of those experiments,
only its upper bound is known until now. 
Since the $\Theta^+(1540)$ has positive strangeness, its minimal quark
content is $(uudd\bar{s})$ and so an exotic state.
Furthermore, the NA49 Collaboration at CERN reported an evidence for the
existence of another exotic $\Xi$ baryon with a mass of 1.862 GeV,
whose isospin is expected to be 3/2 \cite{NA49-03}.

\begin{table}[b]
\centering
\begin{tabular}{c|c|c|c} \hline\hline
Basic reaction & $M(\Theta^+)$ in MeV & $\Gamma(\Theta^+)$ in
MeV & Collaboration/Reference \\ \hline
$\gamma n \to K^+ K^- n $ & $1540 \pm 10$ & $\le 25$ & LEPS \cite{LEPS03} \\
$K^+ \mbox{Xe} \to K^0 p \mbox{Xe}'$ & $1539 \pm 2$ & $\le 9$ &
DIANA \cite{DIANA03} \\
$\gamma d \to K^+ K^- p n$ & $1542 \pm 5$ & $\le 21$ & CLAS \cite{CLAS03-b} \\
$\gamma p \to K^+ K_S^0 n$ & $1540 \pm 4 \pm 2$ & $\le 25$ & SAPHIR
\cite{SAPHIR03} \\
$\gamma p \to \pi^+ K^- K^+ n$ & $1537 \pm 10$ & $\le 31$ & CLAS
\cite{CLAS03-c} \\
$\nu_\mu (\overline{\nu}_\mu) + A \to \mu^- (\mu^+) p K_S^0 X$ & $1533
\pm 5$ & $\le 20$ & BBCN \cite{ADK03} \\
$\gamma p \to \pi^+ K^- K^+ n$ & $1555 \pm 10$ & $\le 26$ & CLAS
\cite{CLAS03-d} \\
$ed \to pK_s^0 X$ & $1528 \pm 2.6 \pm 2.1$ & $13 \pm 9 \pm 3$ & HERMES
\cite{HERMES03} \\
\hline\hline
\end{tabular}
\caption{Summary of the experimental data for the $\Theta^+(1540)$ baryon.}
\label{tab:expt}
\end{table}

\begin{figure}
\centering
\epsfig{file=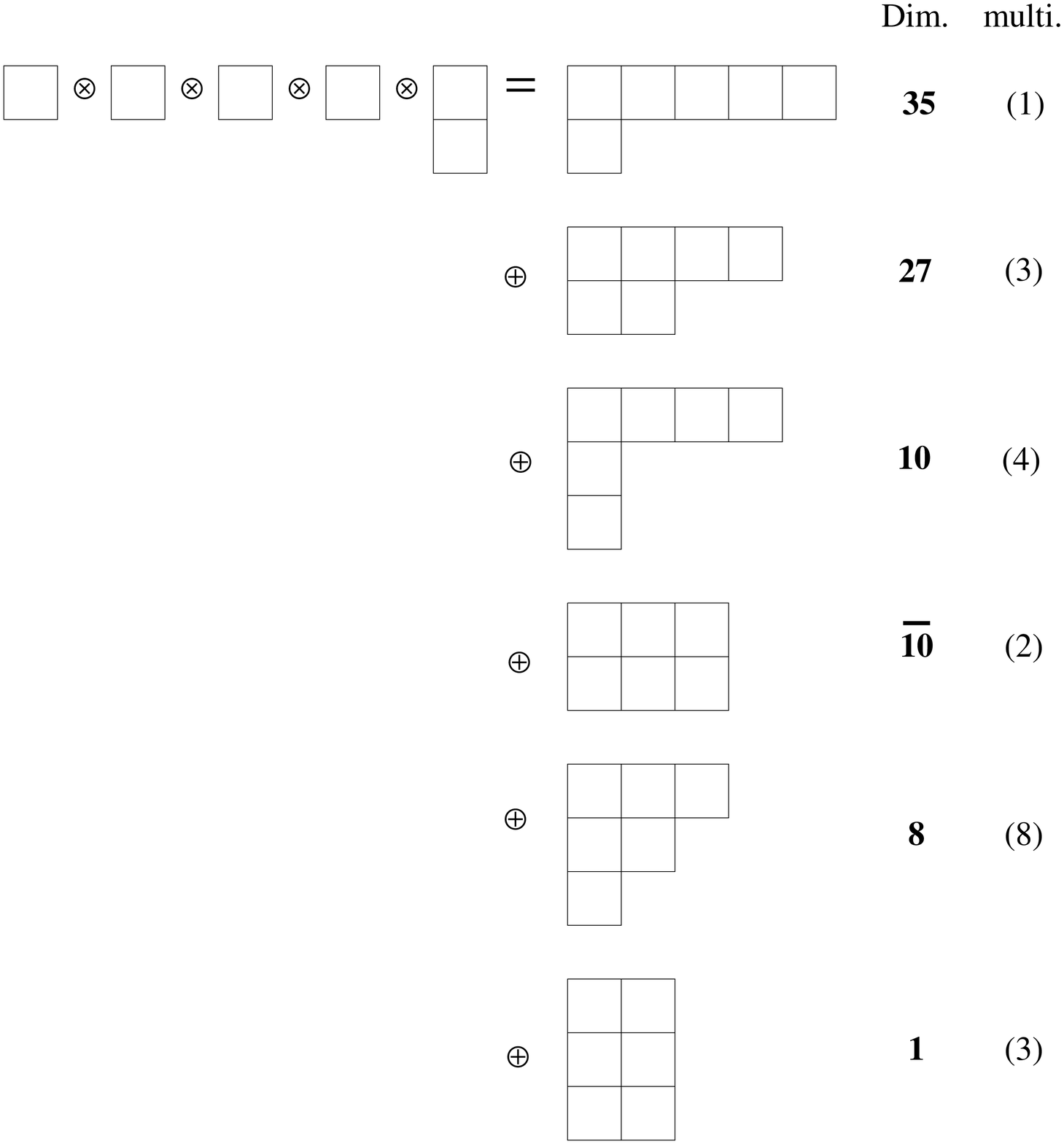, width=0.5\hsize}
\caption{Pentaquark baryons in the quark model. The numbers in
parentheses are the multiplicity.}
\label{fig:penta}
\end{figure}

Theoretically, the chiral soliton model of Ref.~\cite{DPP97} predicted a
narrow pentaquark $\Theta^+$ baryon at a mass of 1.53 GeV, which forms the
baryon antidecuplet with the exotic $\Xi$ particle.
Such states are also anticipated in SU(3) Skyrme models, which allow
all possible SU(3) representation of baryons
\cite{Man84-Chem85-Pras87,Weig98}.%
\footnote{Therefore, the SU(3) Skyrme model contains not only the
pentaquarks but the heptaquarks and etc.}
Although there is no quark degrees of freedom in soliton models, the
positive strangeness of the $\Theta^+$ implies the existence of a strange
anti-quark in its wavefunction and hence the $\Theta^+$ is interpreted
as a pentaquark state in the quark model description.
Exotic hadrons have been a challenge for hadron models and they are expected
to widen our understanding of the hadron structure \cite{Clo03}.
Viewing pentaquark baryons containing four quarks and one anti-quark,
one finds that such baryons can form singlet, octet, decuplet,
antidecuplet, 27-plet, and 35-plet, namely,
\begin{equation}
\bm{3} \otimes \bm{3} \otimes \bm{3} \otimes \bm{3} \otimes
\overline{\bm{3}} = \bm{35} + (3)\bm{27} + (2)\overline{\bm{10}} +
(4)\bm{10} + (8)\bm{8} + (3)\bm{1},
\end{equation}
as shown in Fig.~\ref{fig:penta}, where the numbers in parentheses are
the number of multiplicity.
Then the $\Theta$ baryon that carries hypercharge $Y=+2$ can be a member
of $\overline{\bm{10}}$, $\bm{27}$, or $\bm{35}$.
However, depending on the multiplets, it has different isospin, namely,
the antidecuplet contains isoscalar ($I=0$) $\Theta$, while the $27$-plet
and $35$-plet contain the isovector ($I=1$) $\Theta_1$ and isotensor
($I=2$) $\Theta_2$ particles, respectively.
Thus, the isospin measurement can answer the question in which multiplet
the $\Theta^+(1540)$ resides.
In this respect, the SAPHIR experiment \cite{SAPHIR03} and the HERMES
experiment \cite{HERMES03}, which report no evidence for a signal of the
$\Theta^{++}(1540)$ in the $K^+ p$ invariant mass distribution in the
$\gamma p$ and $ed$ reactions, strongly support that the $\Theta^+(1540)$
is an isosinglet and hence it belongs to antidecuplet, which is also
supported by the $KN$ scattering data analyses of Ref.~\cite{ASW03-ASW03b}.
The low mass of the exotic $\Xi$ states \cite{NA49-03} also seems to
support the antidecuplet nature of the $\Theta^+(1540)$ and
$\Xi(1862)$. (See also Ref.~\cite{FW04} for a critical discussion about
$\Xi(1862)$.)

However, the properties of the $\Theta^+(1540)$, especially its structure
and quantum numbers, are yet to be clarified.
As for the structure of the $\Theta^+$, 
there have been many suggestions in addition to the soliton models.
In Ref.~\cite{KL03a}, Karliner and Lipkin suggested
triquark-diquark model, where the $(ud)$ diquark and the $(ud\bar{s})$
triquark form the $\Theta^+$.
In Ref.~\cite{JW03}, Jaffe and Wilczek advocated diquark-diquark-antiquark
model so that the $\Theta^+$ is a system of $(ud)$-$(ud)$-$\bar{s}$.
In this model, they also considered the mixing of the pentaquark
antidecuplet with the pentaquark octet, which makes it different from the
SU(3) soliton models where the octet describes the normal baryon octet.
Assuming that the nucleon and $\Sigma$ analogues are in the ideal mixing
of the octet and antidecuplet, the nucleon analogue is then identified as
the Roper resonance $N(1440)$.%
\footnote{In Ref.~\cite{OKL03b}, it was pointed out that the Roper
resonance $N(1710)$ cannot be a {\it pure antidecuplet state\/} due to
SU(3) symmetric interactions and $U$-spin conservation. The antidecuplet
cannot couple to decuplet and meson octet while $N(1710)$ has a large
branching ratio into the $\pi\Delta$ channel. Therefore, mixing with other
multiplets is required to identify the $N(1710)$ as a pentaquark
crypto-exotic state. However recent studies on the ideal mixing
of antidecuplet with pentaquark octet show that the ideally mixed state
still has vanishing coupling with the $\pi\Delta$ channel \cite{CD04,LKO04}.}
In these correlated quark configurations, the color-flavor-spin
wavefunction analyses favor the even-parity of the $\Theta^+(1540)$,
which is consistent with the soliton model prediction.
More discussions on the quark model predictions based on the diquark
picture can be found, e.g., in Refs.~\cite{OKL03b,DP03b,SZ03}.
Predictions on the antidecuplet spectrum in various quark models can be
found, e.g., in Refs.~\cite{BGS03,SR03,CCKN03a,CCKN03b,Gloz03a,GK03b}.

Other theoretical investigations on the structure of $\Theta^+$ and
baryon antidecuplet include soliton models
\cite{WK03,Pras03,IKOR03,JM03,BFK03}, 
QCD sum rules \cite{Zhu03,MNNRL03,SDO03},
large $N_c$ QCD \cite{Cohen03-CL03}, and
lattice calculation \cite{CFKK03-Sasaki03}.
Another approach is to explain the $\Theta^+$ as a $KN$ or $K\pi N$
bound state \cite{BM03-KK03-KS03}.
Therefore, understanding the $\Theta^+$ properties is essential to test
those models.
The production of $\Theta^+$ can also be investigated in heavy-ion
collisions as discussed in Ref.~\cite{Rand03-CGKLL03-LTSR03}.
In this case, the $\Theta^+$ yield is expected to carry informations on
the early stage of the heavy ion collisions because of relatively weak
interactions of $\Theta^+$ with the hadrons produced in the collisions.
But the results would be dependent on the detailed structure of
$\Theta^+$.

Among the properties of $\Theta^+(1540)$, the parity quantum number is one
of the important issues to be settled.
If one assumes that all the quarks of the $\Theta^+$ are in the $S$ wave
ground state, it is natural to expect the odd-parity for $\Theta^+$
\cite{CCKN03a,Zhu03,SDO03,CFKK03-Sasaki03}.
However, if the quarks are strongly correlated, the $P$ wave state is
expected to have a lower energy so that the observed $\Theta^+$ has
even-parity \cite{KL03a,JW03}, which is consistent with the soliton model
prediction \cite{DPP97}.
This is also consistent with the quark potential model calculations
\cite{CCKN03b}, where the flavor-spin interaction plays an important role
to give a lower mass to the $P$ wave state.
In this respect, it is interesting to recall the models for heavy
pentaquarks, where an issue is whether the non-strange heavy pentaquarks
like $\Theta_c^0$ and $\Theta_b^+$ are stable against strong decays
\cite{Lip87-GSR87}.
In the bound state model of Skyrmion \cite{CK85-OMRS91-RS93}, such heavy
pentaquarks are described as soliton--heavy-meson bound states.
Although the $\Theta^+$ cannot be a bound state of $KN$ \cite{IKOR03}, a
system like $\overline{D}N$ or $BN$ can form bound states due to the heavy
mass of $D$ and $B$ mesons.
Then the pentaquark masses are lower than the thresholds so their decays by
strong interactions are not allowed energetically.
The detailed model calculation can be found in
Ref.~\cite{OPM94b-OPM94c-OP95}, where the heavy vector mesons are
introduced to satisfy the heavy quark symmetry.
The results show that the relative $P$ wave state is the ground state so
that the lowest $\Theta_c^0$ and $\Theta_b^+$ states have even-parity.
Therefore one would expect even-parity $\Theta^+$ unless level inversion
occurs by the lighter mass of the kaon.

Nevertheless, the parity of the $\Theta^+$ should be determined unambiguously
by experiments through its production and decay processes.%
\footnote{The spin-parity of the $\Theta^+$ can be measured kinematically
from the distribution of its decay into $KN$ if the polarization of
the outgoing nucleon is detected, which is, however, very 
difficult at present.}
The production processes of the $\Theta^+$ baryon have been investigated
by several groups.
In Refs.~\cite{LK03a,LK03b}, the total cross sections of $\Theta^+$
production in photon-nucleon, meson-nucleon, and nucleon-nucleon
reactions were discussed and obtained.
It is then improved by including the tensor coupling terms for
photon-nucleon interaction \cite{NHK03} and $K^*$ exchanges
\cite{OKL03a}.
The obtained results show that the cross sections for the odd-parity
$\Theta^+$ are much smaller than those for the even-parity.
Based on the SU(3) symmetric Lagrangian of Ref.~\cite{OKL03b}, the
$\Theta^+$ production processes in the $KN$ and $NN$ reactions were also
discussed in Ref.~\cite{OKL03c} and the cross section of the $\gamma p \to
\pi^+ K^- \Theta^+$ reaction was obtained in Ref.~\cite{LKK03}.
In order to determine the parity of the $\Theta^+$, various polarization
observables in the $K^+ p \to \pi^+ K^+ n$ reaction \cite{HHO03}, $\gamma
n \to K^- \Theta^+$ \cite{ZA03}, $\gamma n \to K^- K^+ n$ \cite{NT03},
and $ p+ p \to \Sigma + \Theta$ \cite{THH03} reactions have been suggested
and shown to be sensitive to the parity of
the $\Theta^+$.
As a continuation of our works on the $\Theta^+$ production processes
\cite{OKL03a,OKL03c}, we investigate $\gamma N \to \overline{K}^* \Theta^+$
in this paper.
In particular, we investigate the production mechanisms
and various spin asymmetries of the $\gamma n \to K^{*-}
\Theta^+$ and $\gamma p \to \bar{K}^{*0} \Theta^+$ reactions, which can
be experimentally studied, e.g., at Thomas Jefferson National
Accelerator Facility and SPring-8.
By doing so, we also study the dependence of the spin observables on the
parity of the $\Theta^+$.
To be consistent with previous efforts  
for the $\Theta$ production in the literature,
we consider in this work only the elementary processes containing 
the well-known resonances in the intermediate state.
Further development including higher resonances will be
necessary in future though the calculation is limited by
their unknown interactions with the $\Theta$. Nevertheless, 
we discuss briefly how higher resonances can affect our results.

This paper is organized as follows.
In Sec. II, we discuss the effective Lagrangians for the $\gamma + N \to
\overline{K}^* + \Theta^+$ reactions with $(N=p,n)$.
The coupling constants and the production amplitudes are then computed
depending on the parity of the $\Theta^+$.
Section III shows the results for cross sections and various spin
asymmetries, and their physical interpretations are discussed.
The results are summarized in Sec. IV.

\section{Production mechanisms}

\begin{figure}
\centering
\epsfig{file=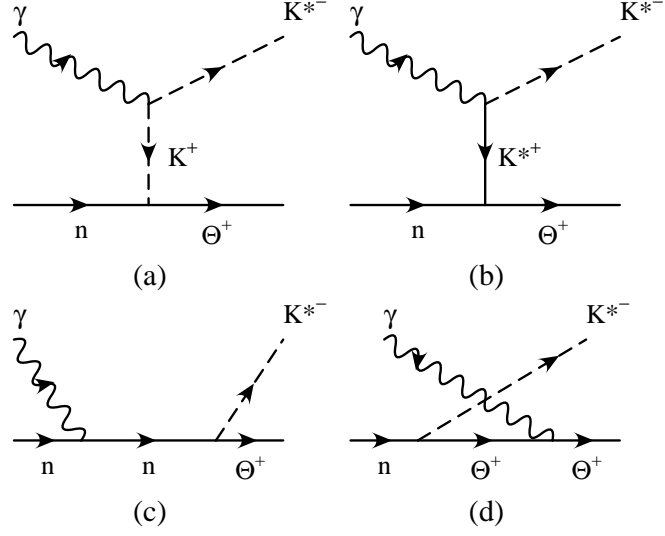, width=0.55\hsize}
\caption{Feynman diagrams for $\gamma n \to K^{*-} \Theta^+$.}
\label{fig:diag1}
\end{figure}

\begin{figure}
\centering
\epsfig{file=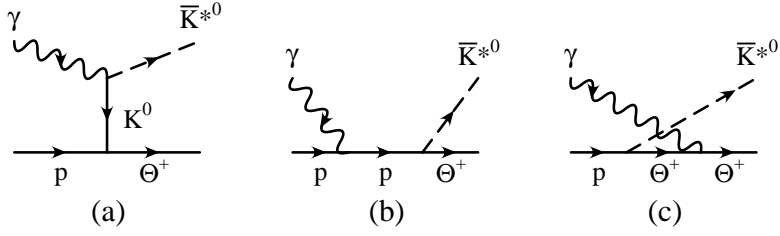, width=0.65\hsize}
\caption{Feynman diagrams for $\gamma p \to \bar{K}^{*0} \Theta^+$.}
\label{fig:diag2}
\end{figure}

Throughout this paper, we assume that the $\Theta^+(1540)$ has the 
quantum numbers, $J=1/2$ and $I=0$, and belongs to baryon antidecuplet.%
\footnote{If it has isospin 2 \cite{CPR03}, it would have very different
production mechanisms.}
We give the results depending on the parity of the $\Theta^+$ so that
various spin asymmetries can be used to study not only the production
mechanisms but also the parity of the $\Theta^+(1540)$.
The reactions that are investigated in this paper are
\begin{equation}
\gamma + n \to K^{*-} + \Theta^+ \qquad \mbox{and} \qquad
\gamma + p \to \bar{K}^{*0} + \Theta^+.
\end{equation}
For the production mechanisms, we consider the tree diagrams as shown in
Figs.~\ref{fig:diag1} and \ref{fig:diag2}.
As in Refs.~\cite{OKL03a,OKL03c}, we consider the lowest baryons for the
intermediate state of the $s$- and $u$-channel diagrams.
Thus, we do not consider the nucleon resonances in the $s$-channel diagrams. 
Note that the $\Delta$ resonances in the intermediate
state are not allowed by isospin.
Similarly, isovector $\Theta_1$ can contribute through the $u$-channel
diagrams but its existence is still not established, so its contribution
will not be considered in this study.
Isotensor $\Theta^{**}$ cannot contribute by isospin conservation.
In $t$-channel diagrams, we consider the $K$ and $K^*$ exchanges.
But we do not consider the exchanges of higher $K$ meson resonances
like axial-vector $K_1(1270)$ meson.
Although its radiative decay width into $K$ meson is measured recently
\cite{KTeV02}, there is no information to constrain its couplings to the
nucleon and the $\Theta^+$ as well as its decay into $K^* \gamma$.
The role of such heavier meson exchanges can be studied when 
experimental data for various $\Theta^+$ production processes
are precise enough.
It should be also noted that the $t$-channel diagram for $\gamma + p \to
\bar{K}^{*0} + \Theta^+$ contains the $K$ exchange only, while $\gamma +
n \to K^{*-} + \Theta^+$ contains $K$ and $K^*$ exchanges
in $t$ channel.

\begin{figure}
\centering
\epsfig{file=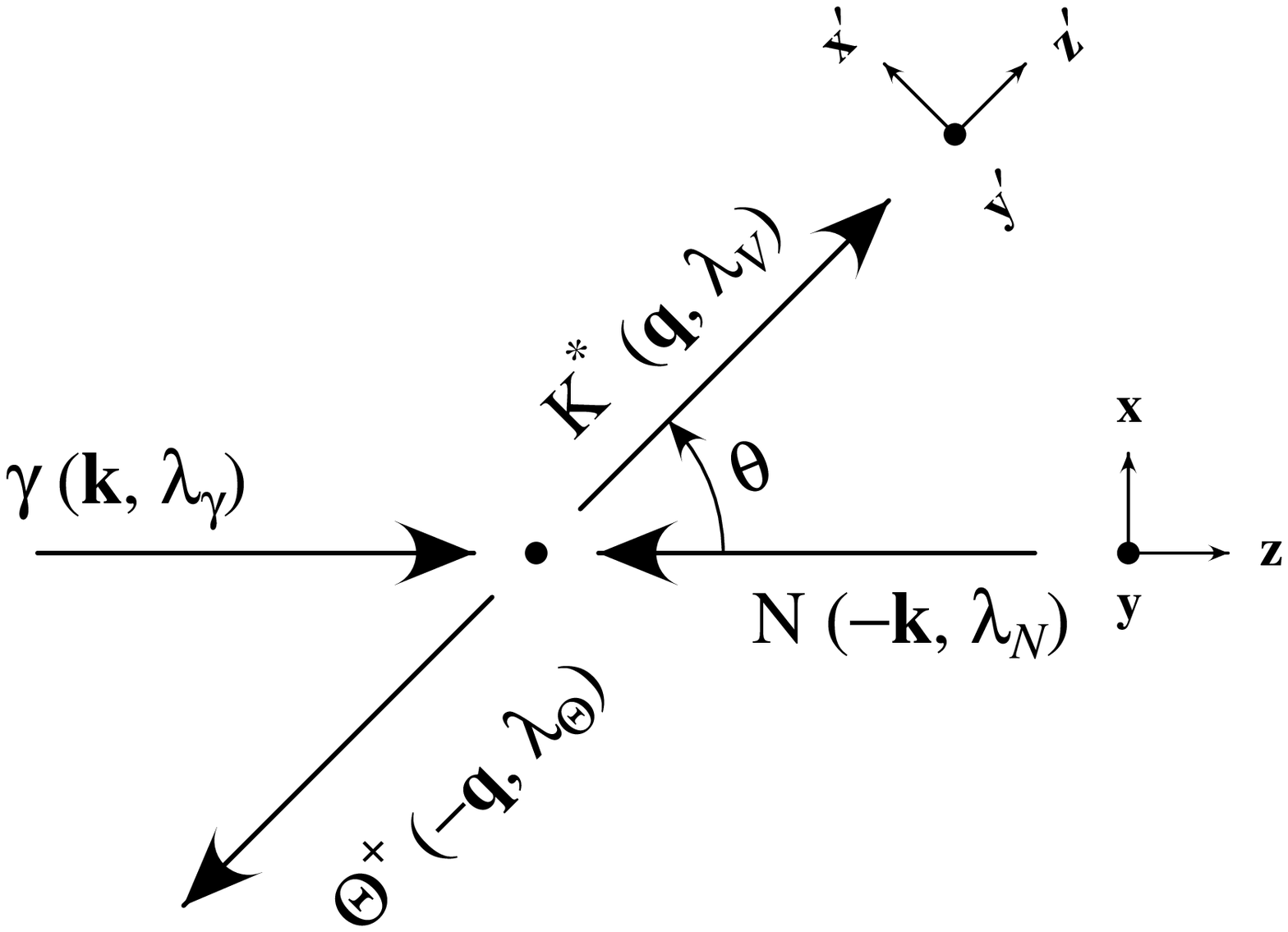, width=0.45\hsize}
\caption{The coordinate system and kinematic variables in the $\gamma + N
\to \overline{K}^* + \Theta$ reaction.}
\label{fig:CM}
\end{figure}

The momenta of the incoming photon, the nucleon, the outgoing
$\overline{K}^*$, and the $\Theta^+$ are $k$, $p$, $q$, and $p'$,
respectively.
The Mandelstam variables are $s=(k+p)^2$, $t=(k-q)^2$ and $u=(p-q)^2$.
In the c.m. frame, the momenta are given by $k = (\nu, {\bf k})$, $p
= (E_N^{}, -{\bf k})$, $q = (E_V^{}, {\bf q})$ and $p' = (E_\Theta^{},
-{\bf q})$ as shown in Fig.~\ref{fig:CM}, which also defines the
scattering angle $\theta$ and the helicities of the particles.

The effective Lagrangians containing $\Theta^+(1540)$ are obtained
from the SU(3) symmetric Lagrangian for the interactions of the baryon
antidecuplet with the meson octet and baryon octet \cite{OKL03b},
\begin{equation}
\mathcal{L}_{\overline{D}PB} = g \overline{T}_{jkl} 
P^j_m B^k_n \epsilon^{lmn} + \mbox{H.c.},
\end{equation}
where $T^{ijk}$ is the baryon antidecuplet, $P^j_m$ the
pseudoscalar/vector meson octet and $B^k_n$ the baryon octet.
With the proper Lorentz structure of the interactions, this leads to
\begin{eqnarray}
\mathcal{L}_{KN\Theta}^\pm &=& g_{KN\Theta}^{}
 \overline{\Theta} \Gamma^\pm \overline{K}^c N
+ \mbox{ H.c.},
\label{KNT}
\\
\mathcal{L}_{K^*N\Theta}^\pm &=& g_{K^*N\Theta}^{} \overline{\Theta} \left(
\Gamma_\mu^\pm \overline{K}^{*c\mu} - \frac{\kappa^T_{K^*N\Theta}}
{M_N + M_\Theta} \Gamma^\mp
\sigma^{\mu\nu} \partial_\nu \overline{K}_\mu^{*c} \right) N
+ \mbox{ H.c.},
\end{eqnarray}
where
\begin{equation}
\left( \begin{array}{c} \Gamma^+ \\ \Gamma^- \end{array}
\right)
= \left( \begin{array}{c} i \gamma_5 \\ 1 \end{array}
\right), \qquad
\left( \begin{array}{c} \Gamma_\mu^+ \\ \Gamma_\mu^- \end{array}
\right)
= \left( \begin{array}{c} \gamma_\mu \\ i
\gamma_5 \gamma_\mu \end{array} \right),
\end{equation}
and
\begin{equation}
N = \left( \begin{array}{c} p \\ n \end{array} \right), \qquad
K^c = \left( \begin{array}{c} \bar{K}^0 \\ -K^- \end{array} \right),
\qquad
K^{*c} = \left( \begin{array}{c} \bar{K}^{*0} \\ -K^{*-} \end{array} \right).
\end{equation}
The upper (lower) sign of $\mathcal{L}_{KN\Theta}^\pm$ and
$\mathcal{L}_{K^*N\Theta}^\pm$ is for the even (odd) parity $\Theta^+$.
The Lagrangian of the $\gamma\Theta\Theta$ interaction is given by 
\begin{eqnarray}
\mathcal{L}_{\Theta\Theta\gamma} &=& -e \overline{\Theta} \left[ A_\mu
\gamma^\mu - \frac{\kappa_\Theta^{}}{2M_\Theta} \sigma_{\mu\nu}
\partial^\nu A^\mu \right] \Theta,
\end{eqnarray}
where $\kappa_\Theta^{}$ is the anomalous magnetic moment of $\Theta^+$.
Other interactions needed to compute the diagrams of
Figs.~\ref{fig:diag1} and \ref{fig:diag2} are
\begin{eqnarray}
\mathcal{L}_{NN\gamma} &=& -e \overline{N} \left[ A_\mu \gamma^\mu
\frac{1+\tau_3}{2} - \frac{1}{4M_N} \left\{ \kappa_p + \kappa_n + \tau_3
(\kappa_p - \kappa_n) \right\} \sigma_{\mu\nu} \partial^\nu A^\mu
\right] N,
\\
\mathcal{L}_{K^*K\gamma} &=& g_{K^*K\gamma}^{}
\varepsilon^{\mu\nu\alpha\beta} \partial_\mu A_\nu \left(
\partial_\alpha K_\beta^{*-} K^+ + \partial_\alpha \bar{K}_\beta^{*0} K^0
\right) + \mbox{ H.c.},
\\
\mathcal{L}_{K^* K^*\gamma} &=& - ie A^\mu \left\{ K^{*-\nu}
\left( \partial_\mu K_\nu^{*+} - \partial_\nu K^{*+} \right) - \left(
\partial_\mu K_\nu^{*-} - \partial_\nu K^{*-}_\mu \right) K^{*+\nu}
\right\}.
\end{eqnarray}

The coupling constants are determined as follows.
First, the coupling constant $g_{KN\Theta}^{}$ can be determined by the
decay width of the $\Theta^+(1540)$.
However, only its upper bound is known at present, so we have to rely on
model predictions or analyses of other reactions.
In the chiral soliton model, the decay width $\Gamma(\Theta)$ was
estimated to be 15 MeV \cite{DPP97} or 5 MeV \cite{PSTCG00}.
Recent analyses on $KN$ scattering data favor such a small decay
width of the $\Theta^+$ \cite{HK03-Nuss03-CN03}, but a much smaller
decay width, namely, 1 MeV or even less, is claimed by
Refs.~\cite{ASW03-ASW03b}.
In this study, leaving its determination to future experimental
investigations, we use $\Gamma(\Theta) = 1$ MeV, which gives
\begin{subequations}
\begin{equation}
g_{KN\Theta}^{} = 0.984,
\end{equation}
for the even-parity $\Theta^+$, and
\begin{equation}
g_{KN\Theta}^{} = 0.137,
\end{equation}
\label{gK}
\end{subequations}
for the odd-parity $\Theta^+$.
There is no direct information for $g_{K^*N\Theta}^{}$, but using quark
model wavefunctions for $\Theta^+$ one can estimate the ratio
$g_{K^* N \Theta}^{}/g_{KN\Theta}^{}$, which gives
\begin{subequations}
\begin{equation}
g_{K^* N \Theta}^{}/g_{KN\Theta}^{} = \sqrt3,
\end{equation}
for the even-parity $\Theta^+$ \cite{CD04,CCKN03c}, and
\begin{equation}
g_{K^* N \Theta}^{}/g_{KN\Theta}^{} = 1/\sqrt3,
\end{equation}
\label{gK*}
\end{subequations}
for the odd-parity $\Theta^+$ \cite{CCKN03a}.
In this paper, we use the above values for our numerical calculation.
It also shows that the ratio, $g_{K^* N \Theta}^{}/g_{KN\Theta}^{}$, is
quite dependent of the parity of $\Theta^+$ and hence can be a tool to
probe the $\Theta^+$ parity.
Some qualitative estimate on this ratio can be obtained
by measuring the spin asymmetries in the $\gamma  N \to \overline{K}^* \Theta^+$
reaction as will be discussed below.
As in Ref.~\cite{OKL03a}, we take 
the tensor coupling, $\kappa^T_{K^*N\Theta}=0$ keeping in mind that the
nonvanishing tensor coupling would affect especially the predictions for
spin asymmetries.
The anomalous magnetic moment $\kappa_\Theta^{}$ of the $\Theta^+$, which
should reveal the structure of the $\Theta^+$, depends on the hadron models
\cite{KP03-LHDCZ03}.
But the results of Ref.~\cite{OKL03a} show that the sensitivity to the
$\kappa_\Theta^{}$ term is overwhelmed by the variation of the 
$g_{K^* N\Theta}^{}$ coupling as long as $|\kappa_\Theta^{}| < 1$. 
In this study we use $\kappa_\Theta^{} = -0.9$ assuming that $\mu_\Theta
\sim 0.1$ \cite{KP03-LHDCZ03}.
The anomalous magnetic moments of the nucleon are $\kappa_p^{} = 1.79$ and
$\kappa_n^{} = -1.91$.
For the $K^* K \gamma$ coupling, we use the experimental data for the
$K^*$ radiative decays as in Ref.~\cite{OKL03a}, which gives
$g_{K^* K \gamma}^{0} = -0.388$ GeV$^{-1}$ for the neutral decay and
$g_{K^* K \gamma}^{c} = 0.254$ GeV$^{-1}$ for the charged decay.
Here the phases of $g_{K^* K \gamma}^{}$ are fixed according to the
SU(3) symmetry and vector meson dominance \cite{BBC85}.

Now we consider the production amplitudes of $\gamma + N \to \overline{K}^*
+ \Theta^+$.
The production amplitudes of this reaction are generally written in the
form of
\begin{equation}
\mathcal{M} = \varepsilon^*_\nu(K^*) \bar{u}_\Theta(p')
\mathcal{M}^{\mu\nu} u_p(p) \varepsilon_\mu(\gamma),
\end{equation}
where $\varepsilon$'s are the polarization vectors of the $K^*$ and
photon.
With the effective Lagrangians given above, the production amplitudes
for $\gamma + n \to K^{*-} + \Theta^+$ shown in
Fig.~\ref{fig:diag1} read
\begin{subequations}
\begin{eqnarray}
\mathcal{M}^{\mu\nu}_{(a)} &=& \frac{g^c_{K^*K\gamma} g_{KN\Theta}^{}}{
t - M_K^2} \varepsilon^{\mu\nu\alpha\beta} k_\alpha q_\beta
\Gamma^\pm  \,F_{2(a)}(s,t,u), 
\\
\mathcal{M}^{\mu\nu}_{(b)} &=& \frac{e g_{K^*N\Theta}^{}}{t - 
M_{K^*}^2} \left( q^\alpha g^{\mu\nu} - k^\nu g^{\mu\alpha} - 2 q^\mu
g^{\nu\alpha} \right) \left\{ g_{\alpha\beta} - \frac{1}{M_{K^*}^2}
(k-q)_\alpha (k-q)_\beta \right\}
\nonumber \\ && \mbox{} \times
\left\{ \Gamma^{\pm\beta} + \frac{i\kappa^T_{K^*N\Theta}}{M_N +
M_\theta} \Gamma^\mp \sigma^{\beta\delta} (k-q)_\delta \right\}
\,F_{2(b)}(s,t,u),
\\
\mathcal{M}^{\mu\nu}_{(c)} &=& \frac{i e g_{K^*N\Theta}^{}}{s -
M_N^2} \frac{\kappa_n}{2M_N} \left( \Gamma^{\pm \nu} - \frac{i
\kappa^T_{K^*N\Theta}}{M_N + M_\Theta} \Gamma^\mp \sigma^{\nu\alpha}
q_\alpha \right) (k\!\!\!/ + p\!\!\!/ + M_N ) \sigma^{\mu\beta} k_\beta
\nonumber \\ && \mbox{} \times
\,F_{2(c)}(s,t,u),
\\
\mathcal{M}^{\mu\nu}_{(d)} &=& \frac{eg_{K^*N\Theta}^{}}{u -
M_\Theta^2} \left[ \gamma^\mu + \frac{i \kappa_\Theta^{}}{2M_\Theta}
\sigma^{\mu\alpha} k_\alpha \right] (p\!\!\!/ - q\!\!\!/ + M_\Theta)
\nonumber \\ && \mbox{} \times
\left( \Gamma^{\pm\nu} - \frac{i\kappa^T_{K^*N\Theta}}{M_N + M_\Theta}
\Gamma^\mp \sigma^{\nu\beta} q_\beta \right)  \,F_{2(d)}(s,t,u),
\end{eqnarray}
\label{gam-n}
\end{subequations}
where $g^c_{K^*K\gamma} = 0.254$ GeV$^{-1}$ is obtained from
$\Gamma( K^{*\pm} \to K^\pm \gamma)$.

For $\gamma + p \to \bar{K}^{*0} \Theta^+$ reaction depicted in
Fig.~\ref{fig:diag2}, we have
\begin{subequations}
\begin{eqnarray}
\mathcal{M}^{\mu\nu}_{(a)} &=& - \frac{g^0_{K^*K\gamma}
g_{KN\Theta}^{}}{t - M_K^2} \varepsilon^{\mu\nu\alpha\beta}
k_\alpha q_\beta \Gamma^\pm  \,F_{3(a)}(s,t,u),
\\
\mathcal{M}^{\mu\nu}_{(b)} &=& - \frac{e g_{K^*N\Theta}^{}}{s -
M_N^2}
\left[ \Gamma^{\pm \nu} - \frac{i \kappa^T_{K^*N\Theta}}{M_N+M_\Theta}
\Gamma^\mp \sigma^{\nu\alpha} q_\alpha \right] ( k\!\!\!/ + p \!\!\!/ +
M_N ) \nonumber \\
&& \mbox{} \times 
\left[ \gamma^\mu + \frac{i \kappa_p}{2M_N} \sigma^{\mu\beta} k
_\beta \right]  \,F_{3(b)}(s,t,u),
\\
\mathcal{M}^{\mu\nu}_{(c)} &=& - \frac{e g_{K^*N\Theta}^{}}{u -
M_\Theta^2} \left[ \gamma^\mu + \frac{i \kappa_\Theta^{}}{2M_\Theta}
\sigma^{\mu\alpha} k_\alpha \right] (p\!\!\!/ - q\!\!\!/ + M_\Theta)
\nonumber \\
&& \mbox{} \times
\left[ \Gamma^{\pm \nu} - \frac{i \kappa^T_{K^*N\Theta}}{M_N+M_\Theta}
\Gamma^\mp \sigma^{\nu\beta} q_\beta \right]  \,F_{3(c)}(s,t,u),
\end{eqnarray}
\label{gam-p}
\end{subequations}
where $g^0_{K^*K\gamma} = -0.388$ GeV$^{-1}$ is obtained from
$\Gamma( K^{*0} \to K^0 \gamma)$.

As in our previous calculations \cite{OKL03a,OKL03c}, we employ the form
factor of the form 
\begin{equation}
F(r,M_{\rm ex}) = \frac{\Lambda^4}{\Lambda^4 + (r - M_{\rm ex}^2)^2},
\label{ff:ours}
\end{equation}
where $M_{\rm ex}$ is the mass of the exchanged particle and $r$ is its
momentum squared.

The amplitudes in Eqs.~(\ref{gam-n}) and (\ref{gam-p}) satisfy the gauge
invariance condition without form factors.
However introducing different form factors at each vertex breaks gauge
invariance.
There are several recipes in restoring gauge invariance with the use of
phenomenological form factors \cite{OKL03a,LKK03}, and
the results depend on the employed form factors and the way to
restore gauge invariance.
In order to keep gauge invariance in a simple way, we use
\begin{eqnarray}
F_{2(b)} = F_{2(d)} = \left\{ F(t,M_{K^*})^2 + F(u,M_\Theta)^2 \right\}/2,
\quad
F_{2(a)} = F(t,M_{K})^2, \quad F_{2(c)} = F(s,M_N)^2,
\nonumber \\
\end{eqnarray}
and
\begin{equation}
F_{3(a)} = F(t,M_{K})^2, \qquad
F_{3(b)} = F_{2(c)} = \left\{ F(s,M_N)^2 + F(u,M_\Theta)^2 \right\}/2.
\end{equation}
This is an unsatisfactory aspect of this hadronic model approach, but it
should be sufficient for this qualitative study.
For the cutoff parameter, we use $\Lambda = 1.8$ GeV as in
Refs.~\cite{OKL03a,OKL03c}.

\section{Results}

Before presenting the results for the $\gamma N \to \overline{K}^* \Theta^+$
reaction, let us discuss the uncertainties arising from neglecting
intermediate baryon resonances in $s$ and $u$ channels.
The production mechanisms include the $s$
and $u$ channel diagrams as shown in Figs.~\ref{fig:diag1}(c,d) and
\ref{fig:diag2}(b,c).
In addition to the ground state resonances that we currently have,
the nucleon resonances ($N^*$) can intermediate the $s$ channel diagrams
and $\Theta$ excitations can contribute to the $u$ channel.  
Those contributions, in principle, should be included as some of 
the physical quantities, especially some spin asymmetries, may be sensitive
to such resonances as in the case of vector meson photoproduction
\cite{OL02}.
At this stage, the possible contribution from $\Theta$ excitations can be
ignored as their existence is not yet confirmed.
As for the $N^*$ contribution, 
if $N^*$ belongs to the baryon antidecuplet, it contributes only to the
neutron target case since it is not allowed for the proton target case due
to $U$ spin symmetry \cite{OKL03b,Clo03}. 
This selection rule does not apply to the $N^*$ belonging to
other multiplets and it can contribute to both targets.
Calculating such contributions, however, should be very limited by 
various unknown couplings.
For example, the $N^*$ contribution in the $s$ channel contains rather
unknown couplings of $\gamma N N^* $ and $\overline{K}^* \Theta N^*$.
Additional uncertainties come from the unfixed width and mass of $N^*$.
Instead of estimating the $N^*$ contribution with such uncertainties,
we include only the lowest and established states as the
intermediate state of $s$ and $u$ channel diagrams.  
Thus, our results should be understood as a first estimate on the spin
asymmetries in the $\gamma N \to \overline{K}^* \Theta$ reaction and
such uncertainties should be kept in mind.%
\footnote{Such uncertainties are also present in other reaction studies
such as $\gamma N \to \overline{K} N$ and $\gamma N \to K^+ K^- N$, while the
$NN \to Y\Theta$ reaction study suffers from the uncertainties arising
from the initial state interactions.}
Since the contributions from the $s$ and $u$ channel diagrmas with
excited states mostly contribute to large scattering angles, we
restrict our investigation to small scattering angle region.

\begin{figure}
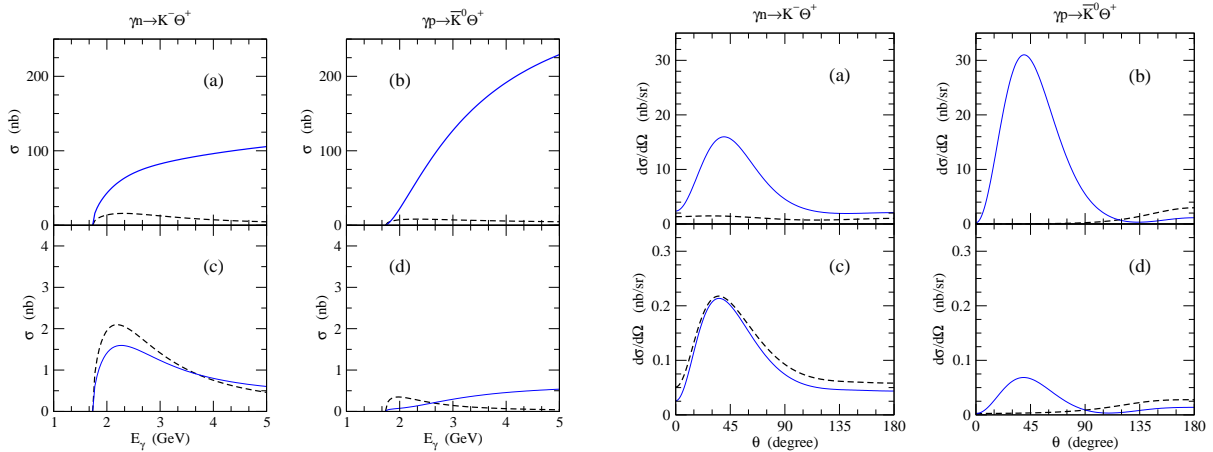

\centering
\epsfig{file=fig5a.eps, width=0.45\hsize} \qquad
\epsfig{file=fig5b.eps, width=0.465\hsize}
\caption{Cross sections for $\gamma + N \to \overline{K} + \Theta^+$.
Total cross sections (left panel) and differential cross
sections at $E_\gamma = 3.0$ GeV (right panel).
(a,c) for $\gamma + n \to K^- + \Theta^+$ and (b,d) for $\gamma + p \to
\bar{K}^0 + \Theta^+$. In (a,b), even-parity is assumed for
$\Theta^+$, and in (c,d), odd-parity of $\Theta^+$ is assumed.
The dashed lines are obtained with $g_{K^* N \Theta}^{} = 0$, and the
solid lines are with $g_{K^* N \Theta}^{}/g_{KN\Theta}^{} =
\sqrt3$ ($1/\sqrt3$) for positive (negative) parity $\Theta^+$
as in Eq.~(\ref{gK*}).}
\label{fig:KT}
\end{figure}

We now present the results for various physical
quantities in the $\gamma + N \to \overline{K}^* + \Theta^+$ reaction based
on the formalism given in the previous section.
For this calculation, we use the couplings $g_{KN\Theta}^{}$ and
$g_{K*N\Theta}^{}$ as in Eqs.~(\ref{gK}) and (\ref{gK*}).
Since the coupling constants are different from those of
Ref.~\cite{OKL03a}, we start with presenting the total cross sections and
differential cross sections at $E_\gamma = 3$ GeV for
$\gamma + N \to \overline{K} + \Theta^+$ in Fig.~\ref{fig:KT} obtained with 
the parameters given above.
Shown in the left panel are the total cross sections and the right panel
presents the results for the differential cross sections.
The dashed lines are obtained with $g_{K^* N \Theta}^{}/g_{KN\Theta}^{}
= 0$ and the solid lines are with the values of Eq.~(\ref{gK*}).
The difference with the results of Ref.~\cite{OKL03a} arises from the
different coupling constant $g_{KN\Theta}^{}$ and the different phase in
$g_{K^* K \gamma}^0$.
This shows that the large value of the ratio
$g_{K*N\Theta}^{}/g_{KN\Theta}^{}$ in the case of even-parity $\Theta^+$
leads to the dominance of $K^*$ exchange in $\Theta^+$ production process,
which is characterized by a peak in the forward scattering angles
near $\theta=\pi/4$. 
The contribution from
$u$ and $s$ channels does not have a distinct peak structure.
The $N^*$ contribution, if it is included, is expected to have a nealy
flat structure as it contributes through the $s$-channel only.
Thus, the distinct peak structure observed near $\theta=\pi/4$ is a
characteristic of the $t$-channel exchanges in this model and it would be
interesting to see whether such structure survives within more sophisticated
models with higher resonances in coupled channels.

\subsection{Cross Sections}

\begin{figure}
\centering
\epsfig{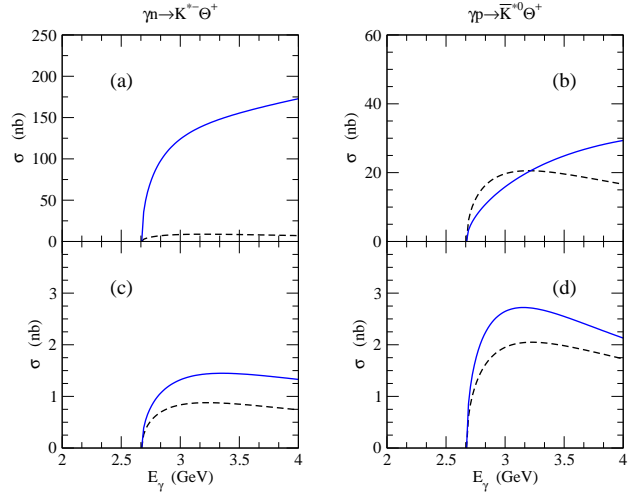}
\caption{Total cross sections for $\gamma + N \to \overline{K}^* + \Theta^+$.
Notations are the same as in Fig.~\ref{fig:KT}.}
\label{fig:tot}
\end{figure}

\begin{figure}
\centering
\epsfig{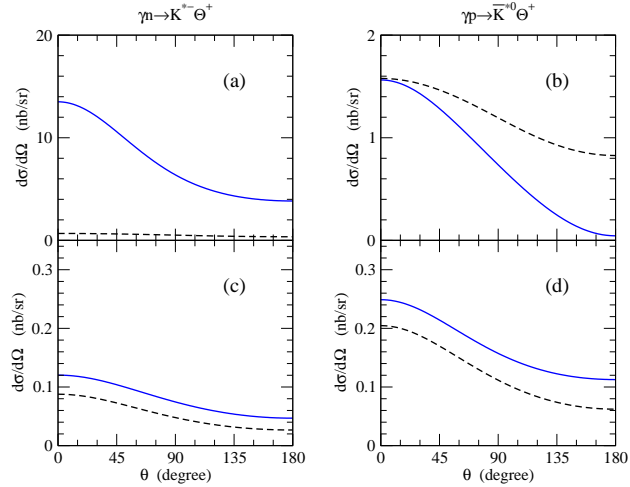}
\caption{Differential cross sections for $\gamma + N \to \overline{K}^*
+ \Theta^+$ at $E_\gamma = 2.8$ GeV.
Notations are the same as in Fig.~\ref{fig:KT}.}
\label{fig:diff}
\end{figure}

In Fig.~\ref{fig:tot}, we present the total cross sections for $\gamma +
N \to \overline{K}^* + \Theta$.
Shown in Fig.~\ref{fig:diff} are the differential cross sections at
$E_\gamma = 2.8$ GeV.
The results show that the cross sections for the odd-parity $\Theta^+$
production are much smaller than those for the even-parity $\Theta^+$
production.
This is because of the smaller coupling constants for the odd-parity
$\Theta^+$.
Figures~\ref{fig:tot} and \ref{fig:diff} also show that the $\gamma p$ 
reaction has larger cross sections than the $\gamma n$ reaction when
$g_{K^*N\Theta}^{} = 0$.
This can be easily understood from the fact that $|g_{K^*K\gamma}^0| >
|g_{K^*K\gamma}^c|$ \cite{BBC85}.
Specifically, without $K^*N\Theta$ interaction, we have the diagrams of
Fig.~\ref{fig:diag1}(a) and \ref{fig:diag2}(a) only.
Therefore, the only difference between the two diagrams lies on the
$K^*K\gamma$ couplings.
Since the $\gamma p$ reaction has neutral kaon exchange, its cross
section is larger than that for the $\gamma n$ reaction.
However, when the $K^*N\Theta$ interaction is turned on, there is 
interference among the production amplitudes so that the cross
sections depend on the magnitude of the $K^*N\Theta$ interaction.
Another interesting observation is that the cross sections for $\gamma +
N \to \overline{K}^* + \Theta^+$ are comparable to or even larger in some
cases than those for the $\gamma + N \to \overline{K} + \Theta^+$ reaction.
But we notice that the shape of the differential cross sections is 
different for $\gamma + N \to \overline{K}^* + \Theta^+$ and
$\gamma + N \to \overline{K} + \Theta^+$.
With nonzero $g_{K^*N\Theta}^{}$ coupling, the $K^*$ production is 
dominantly in forward angles while for the $K$, the
production is peaked near the 45 degrees. 
This is mostly due to the fact that
the role of the $K$ and $K^*$ exchanges is different in these reactions,
e.g., in the $\gamma p$ reaction, the $t$-channel diagram has only $K$
exchange in $\overline{K}^*\Theta^+$ production while in
$\overline{K}\Theta^+$ production the diagram contains only $K^*$ exchange.  
The forward peak in $\gamma + N \to \overline{K}^* + \Theta^+$ 
is expected to persist even when the $N^*$ contribution is
included.  As we have mentioned, the resonance 
contribution in $u$ and $s$-channels
is expected to change the shape in the backward angles. In fact,
this contribution makes the solid curves in Fig.\ref{fig:diff}
flattened in the backward angles.  Thus, the $N^*$ contribution, which
contributes through the $s$-channel, is expected to change the shape
mostly in the backward angles.

\subsection{Single and Double Spin Asymmetries}

\begin{figure}
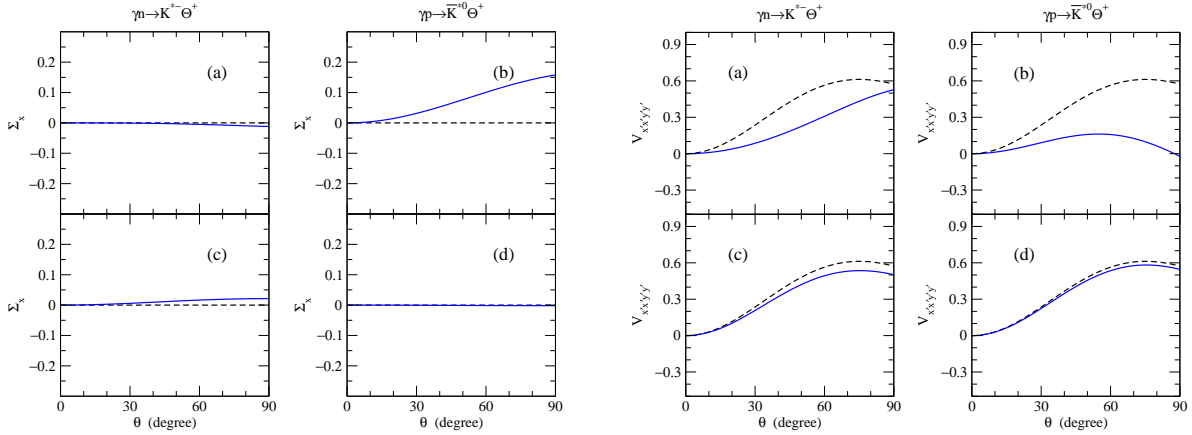

\centering
\epsfig{file=fig8a.eps, width=0.45\hsize} \qquad
\epsfig{file=fig8b.eps, width=0.45\hsize}
\caption{Polarized photon beam asymmetry $\Sigma_x$ (left panel)
and vector meson $(K^*)$ tensor asymmetry $V_{x'x'y'y'}$ (right panel) for
$\gamma + N \to \overline{K}^* + \Theta^+$ at $E_\gamma = 2.8$ GeV.
Notations are the same as in Fig.~\ref{fig:KT}.}
\label{fig:sigma}
\end{figure}

We now compute several single and double spin asymmetries at $E_\gamma =
2.8$ GeV, of which definitions and explicit expressions can be found,
e.g., in Ref.~\cite{TOYM98}.
We first calculate single spin asymmetries: polarized photon beam
asymmetry (analyzing power) $\Sigma_x$ and tensor polarization asymmetry
$V_{x'x'y'y'}$ of the produced $K^*$.
As we have discussed, the contribution from the higher nucleon resonances
would be important in spin asymmetries at large scattering angles.
Therefore, we give our results only for {\it forward\/} scattering angle,
$\theta < 90^\circ$.
The polarized photon beam asymmetry is defined as
\begin{equation}
\Sigma_x = \frac{\sigma^\parallel - \sigma^\perp}{\sigma^\parallel +
 \sigma^\perp},
\end{equation}
where $\sigma^{\parallel}$ ($\sigma^\perp$) is the differential cross section
produced by a photon linearly polarized along the $\hat{\bf x}$ and
($\hat{\bf y}$) axis.
The definition of the tensor polarization asymmetry of the vector meson
can be found, e.g., in Refs.~\cite{TOYM98,KCT98}.
In Fig.~\ref{fig:sigma}, the results for $\Sigma_x$ and $V_{x'x'y'y'}$
are presented.
Here again, the dashed and solid lines are obtained with
$g_{K^*N\Theta}^{}/g_{KN\Theta}^{} = 0$ and $\sqrt3$ ($1/\sqrt3$) for
even (odd) parity of $\Theta^+$, respectively.
Our results show the dependence of the single spin asymmetries on the
production mechanisms.

\begin{figure}
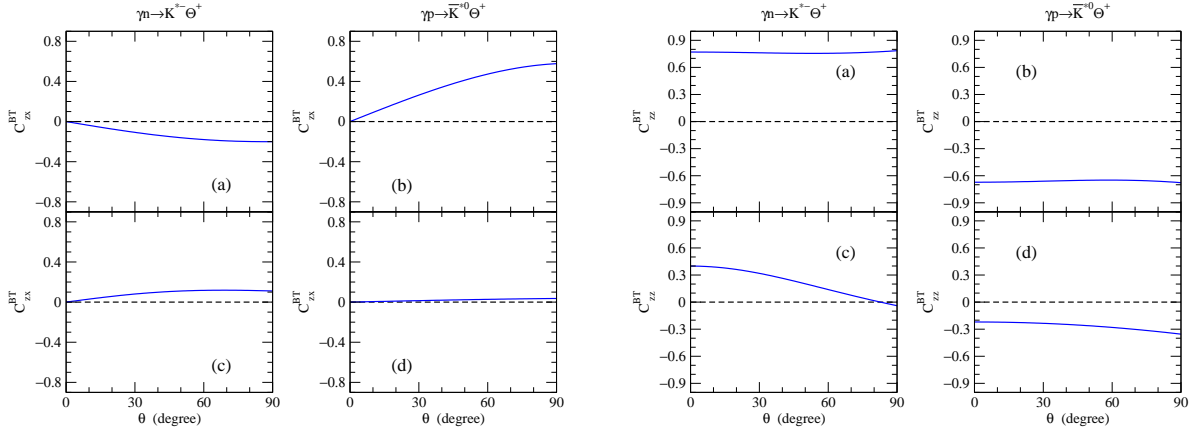

\centering
\epsfig{file=fig9a.eps, width=0.45\hsize} \qquad
\epsfig{file=fig9b.eps, width=0.45\hsize}
\caption{Beam-target double asymmetries $C^{\rm BT}_{zx}$ (left panel) and
$C^{\rm BT}_{zz}$ (right panel) for $\gamma + N \to \overline{K}^* + \Theta^+$
at $E_\gamma = 2.8$ GeV.
Notations are the same as in Fig.~\ref{fig:KT}.}
\label{fig:BT}
\end{figure}

\begin{figure}
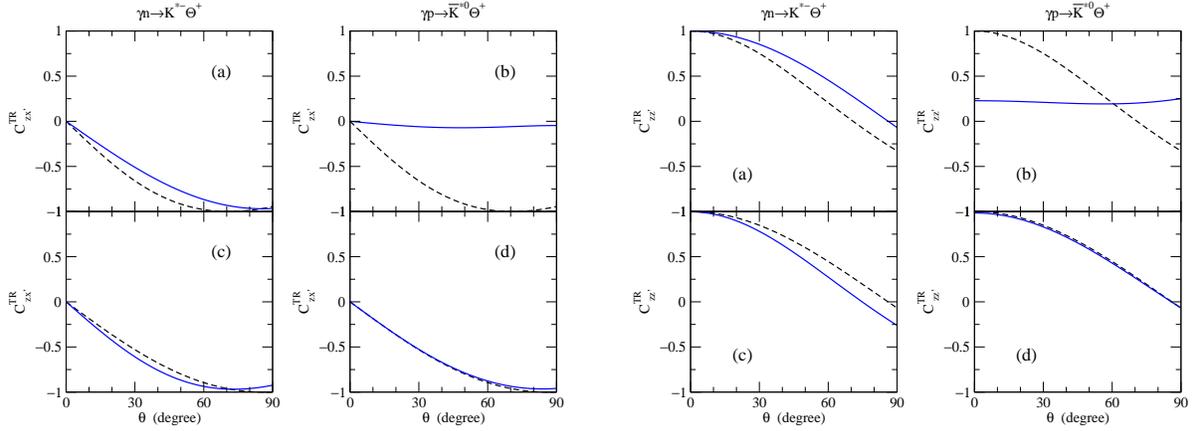

\centering
\epsfig{file=fig10a.eps, width=0.45\hsize} \qquad
\epsfig{file=fig10b.eps, width=0.45\hsize}
\caption{Target-recoil double asymmetries $C^{\rm TR}_{zx'}$ (left panel) and
$C^{\rm TR}_{zz'}$ (right panel) for $\gamma + N \to \overline{K}^*
+ \Theta^+$ at $E_\gamma = 2.8$ GeV.
Notations are the same as in Fig.~\ref{fig:KT}.}
\label{fig:TRz}
\end{figure}

\begin{figure}
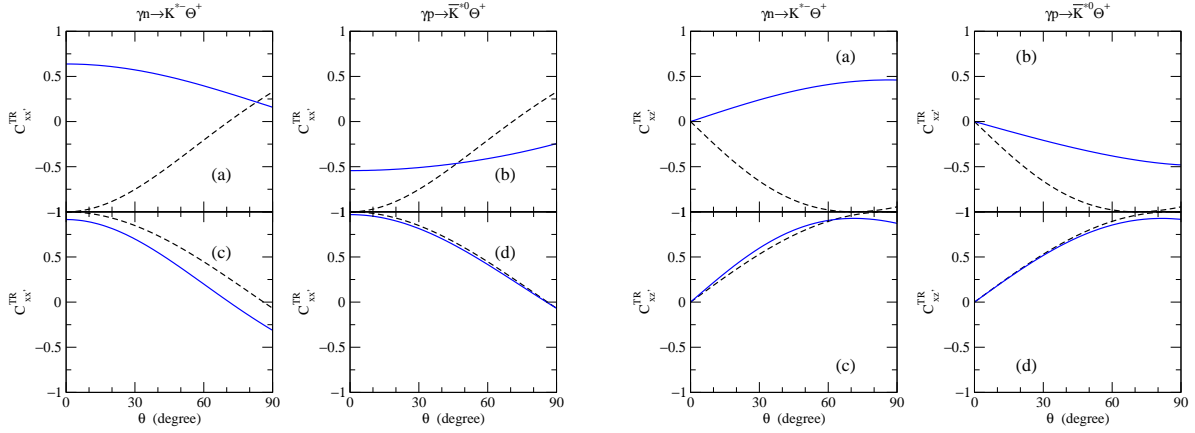

\centering
\epsfig{file=fig11a.eps, width=0.45\hsize} \qquad
\epsfig{file=fig11b.eps, width=0.45\hsize}
\caption{Target-recoil double asymmetries $C^{\rm TR}_{xx'}$ (left panel) and
$C^{\rm TR}_{xz'}$ (right panel) for $\gamma + N \to \overline{K}^*
+ \Theta^+$ at $E_\gamma = 2.8$ GeV.
Notations are the same as in Fig.~\ref{fig:KT}.}
\label{fig:TRx}
\end{figure}

Shown in Figs.~\ref{fig:BT}--\ref{fig:TRx} are the results for
several double spin asymmetries.
Here, we present the results for beam-target (Fig.~\ref{fig:BT}) and
target-recoil (Figs.~\ref{fig:TRz} and \ref{fig:TRx}) asymmetries.
For their definitions, let us consider, e.g., beam-target double
asymmetry $C^{\rm BT}_{zz}$.
The physical meaning of this asymmetry is
\begin{equation}
C^{\rm BT}_{zz} = \frac{\sigma^{z,z} - \sigma^{z,-z}}
{\sigma^{z,z} + \sigma^{z,-z}},
\end{equation}
where the first $z$ means the polarization of the photon beam, i.e., it
is circularly polarized with helicity $+1$, and the second $\pm z$
denotes the direction of the target nucleon polarization.
The reference frame is defined in Fig.~\ref{fig:CM}.
The other double polarizations are defined in a similar way
\cite{TOYM98}.
The obtained results show that the asymmetries depend not only on the
production dynamics but also on the $\Theta^+$ parity.

The most interesting results are the target-recoil double asymmetries
$C^{\rm TR}_{xx'}$ and $C^{\rm TR}_{xz'}$ shown in Fig.~\ref{fig:TRx}.
These asymmetries for the proton targets, in particular, are very
sensitive to the parity of the $\Theta^+$, as their dependence on the
$K^* N \Theta$ interaction is relatively weak because of the absence of
$K^*$ exchange in $t$-channel.
In order to understand the behavior of those asymmetries, let us
consider the helicity amplitudes in the $t$-channel $K$ exchange.
The amplitudes contain
\begin{eqnarray}
\bar{u}_\Theta^{}(p',m_f^{}) \gamma_5 u_N^{}(p,m_i^{}) &\propto&
A_1 (s,t,u) \chi_\Theta^\dagger \bm{\sigma} \cdot \hat{\bf p} \chi_N^{}
- A_2 (s,t,u) \chi_\Theta^\dagger \bm{\sigma} \cdot \hat{\bf p}' \chi_N^{},
\nonumber \\
\bar{u}_\Theta^{}(p',m_f^{}) u_N^{}(p,m_i^{}) &\propto&
B_1 (s,t,u) \chi_\Theta^\dagger \chi_N^{}
- B_2 (s,t,u) \chi_\Theta^\dagger \bm{\sigma} \cdot \hat{\bf p}
\bm{\sigma} \cdot \hat{\bf p}' \chi_N^{},
\end{eqnarray}
where $\chi$'s are the Pauli spinors and $A_i$ and $B_i$ are some
functions of the Mandelstam variables.
Then in the c.m. frame (Fig.~\ref{fig:CM}), $\hat{\bf p} = -\hat{\bf
z}$, and at $\theta = \pi$ so that $\hat{\bf p}' = \hat{\bf z}$,
we have
\begin{eqnarray}
\bar{u}_\Theta^{}(p',m_f^{}) \gamma_5 u_N^{}(p,m_i^{}) &\propto&
- \left[ A_1(s,t,u) + A_2(s,t,u) \right] \chi_\Theta^\dagger \sigma_3
\chi_N^{}
\nonumber \\
&=& - \left[ A_1(s,t,u) + A_2(s,t,u) \right] m_i^{}
\delta_{m_i^{},m_f^{}},
\nonumber \\
\bar{u}_\Theta^{}(p',m_f^{}) u_N^{}(p,m_i^{}) &\propto&
\left[ B_1 (s,t,u) + B_2(s,t,u) \right] \chi_\Theta^\dagger \chi_N^{}
\nonumber \\
&=& \left[ B_1 (s,t,u) + B_2(s,t,u) \right] \delta_{m_i^{},m_f^{}}.
\label{limit}
\end{eqnarray}
Furthermore, since
\begin{eqnarray}
C^{\rm TR}_{xx'} &\propto&
\langle - \textstyle\frac12 | \mathcal{M} | + \textstyle\frac12 \rangle^*
\langle + \textstyle\frac12 | \mathcal{M} | - \textstyle\frac12 \rangle
+
\langle - \textstyle\frac12 | \mathcal{M} | - \textstyle\frac12 \rangle^*
\langle + \textstyle\frac12 | \mathcal{M} | + \textstyle\frac12 \rangle,
\nonumber \\
C^{\rm TR}_{xz'} &\propto&
\langle + \textstyle\frac12 | \mathcal{M} | + \textstyle\frac12 \rangle^*
\langle + \textstyle\frac12 | \mathcal{M} | - \textstyle\frac12 \rangle
-
\langle - \textstyle\frac12 | \mathcal{M} | + \textstyle\frac12 \rangle^*
\langle - \textstyle\frac12 | \mathcal{M} | - \textstyle\frac12 \rangle,
\end{eqnarray}
in the form of $\langle \lambda_f^{} | \mathcal{M} | \lambda_i^{}
\rangle$, where the sum over the other helicities is understood, and using
Eq.~(\ref{limit}) at $\theta \to pi$, we can find that
$C^{\rm TR}_{xx'}$ has different sign depending on the parity of the
$\Theta^+$ and $C^{\rm TR}_{xz'} = 0$ at $\theta = \pi$.
This conclusion holds also at $\theta = 0$.
Thus the results shown in Fig.~\ref{fig:TRx} can be understood in this
kinematic region.
Of course, these results should be modified to some extent by including
$K^*N\Theta$ interactions, but the results of Fig.~\ref{fig:TRx} show that
it does not change so much for the case of the proton targets at least
in the forward scattering region.
However, for the neutron targets, the interference between the two
interactions is large so that the results are dependent on the
$g_{K^*N\Theta}^{}$ coupling constant.
Although the target-recoil double asymmetries are very sensitive to the
parity of the $\Theta^+$, experimental measurements of those asymmetries
would be very hard because of the difficulties with identifying the
helicity of $\Theta^+$.

\subsection{Parity Asymmetry and Photon Asymmetry}

We now consider the parity asymmetry $P_\sigma$ and the photon asymmetry
$\Sigma_V^{}$ in $\gamma + N \to \overline{K}^* + \Theta^+$.
These asymmetries can be measured by observing the decay angular
distribution of the $\bar{K}^*$ vector meson produced by linearly polarized
photon beams \cite{SSW70}.
If we define $\tilde\sigma_\parallel$ and $\tilde\sigma_\perp$ as the
cross sections for symmetric decay particle pairs, i.e., kaon and pion
pairs, produced parallel and normal to the photon polarization plane
respectively, the photon polarization asymmetry $\Sigma_V^{}$ is defined by 
\begin{equation}
\Sigma_V^{} \equiv \frac{\tilde\sigma_\parallel - \tilde\sigma_\perp}
{\tilde\sigma_\parallel + \tilde\sigma_\perp},
\end{equation}
which can be related to the density matrix elements of the $K^*$ vector
meson (when produced by linearly polarized photon beam) as
\begin{equation}
\Sigma_V^{} =
\frac{ \rho^1_{11} + \rho^1_{1-1} }{ \rho^0_{11} + \rho^0_{1-1} }.
\end{equation}
The definitions for the density matrix $\rho^i_{\lambda\lambda'}$ can be
found in Ref.~\cite{SSW70}.

Another interesting quantity is the parity asymmetry $P_\sigma$.
This can be defined by decomposing the helicity amplitudes as
\begin{equation}
\mathcal{M} = \mathcal{M}^N + \mathcal{M}^U,
\end{equation}
or
\begin{equation}
\mathcal{M}^{N/U}_{\lambda_V^{} \lambda_\Theta^{}, \lambda_\gamma^{}
\lambda_N^{}} =
\frac12 \left\{
\mathcal{M}_{\lambda_V^{} \lambda_\Theta^{}, \lambda_\gamma^{} \lambda_N^{}}
\mp
(-1)^{\lambda_V^{}} \mathcal{M}_{-\lambda_V^{} \lambda_\Theta^{},
-\lambda_\gamma^{} \lambda_N^{}}
\right\}.
\end{equation}
This decomposition is from the observation that if only natural parity
[$\eta=(-1)^j$, where $\eta$ and $j$ are the parity and spin of the
exchanged particle] or only unnatural parity [$\eta=-(-1)^j$] exchange in the
$t$-channel contributes, one has an additional symmetry to leading order
in the incoming photon energy \cite{CSM68},
\begin{eqnarray}
\mathcal{M}_{-\lambda_V^{} \lambda_\Theta^{}, -\lambda_\gamma^{}
\lambda_N^{}} &=& \pm (-1)^{\lambda_V^{} - \lambda_\gamma^{}}
\mathcal{M}_{\lambda_V^{} \lambda_\Theta^{}, \lambda_\gamma^{} \lambda_N^{}}
\nonumber \\
&=& \mp (-1)^{\lambda_V^{}}
\mathcal{M}_{\lambda_V^{} \lambda_\Theta^{}, \lambda_\gamma^{} \lambda_N^{}},
\label{addsym}
\end{eqnarray}
where the upper (lower) sign applies to natural (unnatural) parity exchange.
Then one can decompose the cross sections due to natural and
unnatural parity exchanges, i.e., signature of the exchanged particle,
and the parity asymmetry is defined as
\begin{equation}
P_\sigma \equiv \frac{\sigma^N - \sigma^U}{\sigma^N + \sigma^U}
= 2 \rho_{1-1}^1 - \rho^1_{00},
\end{equation}
where $\sigma^N$ and $\sigma^U$ are the contributions of natural and
unnatural parity exchanges to the cross section, and we have written
$P_\sigma$ in terms of the density matrix elements.
So when we have natural parity exchange (like $K^*$ exchange) only, we
get $P_\sigma = +1$ and we expect $P_\sigma = -1$ for unnatural parity
exchange (like $K$ exchange) only.
Note also that the relation (\ref{addsym}) is exact in large energy
limit and that there can be some modifications at low energies.
Nevertheless, this quantity can give some information on the dominant
$t$-channel exchange process.
In helicity conserving processes, the two asymmetries, $\Sigma_V^{}$ and
$P_\sigma$, have similar values as in our case.

\begin{figure}
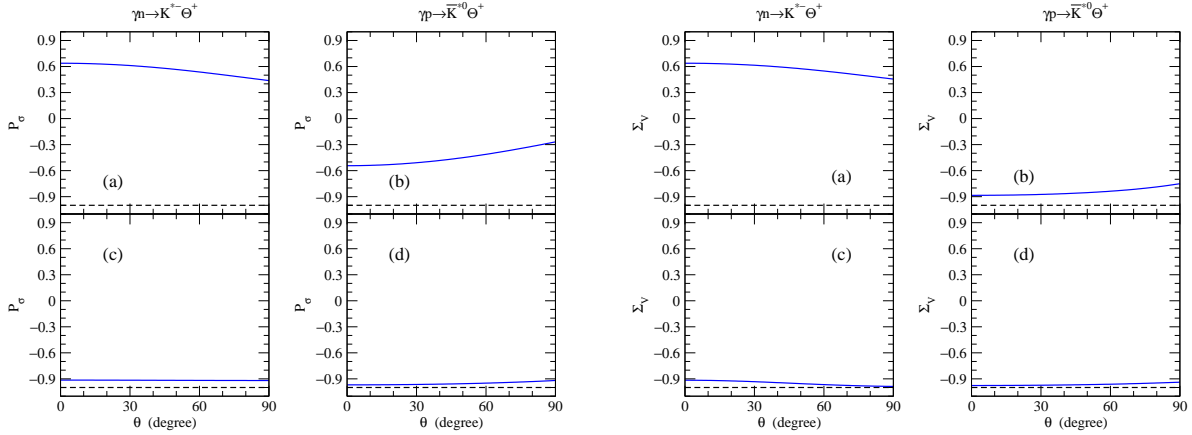

\centering
\epsfig{file=fig12a.eps, width=0.45\hsize} \qquad
\epsfig{file=fig12b.eps, width=0.45\hsize}
\caption{Parity asymmetry $P_\sigma$ (left panel) and photon
polarization asymmetry $\Sigma_V^{}$ (right panel) for
$\gamma + N \to \overline{K}^* \Theta^+$ at $E_\gamma = 2.8$ GeV.
Notations are the same as in Fig.~\ref{fig:KT}.}
\label{fig:pol}
\end{figure}

Shown in Fig.~\ref{fig:pol} are our predictions for the asymmetries
$P_\sigma$ and $\Sigma_V^{}$.
Since they are very sensitive to the spin-parity of the exchanged
particle, we can extract informations on the nature of the production
mechanisms.
For example, if we turn off $K^* N \Theta$ interaction by setting
$g_{K^*N\Theta}^{} = 0$, we have only $t$-channel $K$ exchange.
Therefore, we expect $P_\sigma = -1$ in this case for both $\gamma
n$ and $\gamma p$ reactions.
When $g_{K^*N\Theta}^{} \neq 0$, we get $t$-channel $K^*$ exchange in
the $\gamma n$ reaction, which is, however, not allowed in the $\gamma
p$ reaction.
In both cases, of course, the $s$ and $u$ channel diagrams can 
make $P_\sigma$ deviate from $\pm 1$, which however does not change the
sign of $P_\sigma$ in the forward scattering angles.
Thus we expect that $P_\sigma$ varies between $-1$ and $+1$ depending on
the relative size of the $K$ and $K^*$ exchanges in the $\gamma n$
reaction, while $P_\sigma$ is close to $-1$ in the $\gamma p$ reaction
at the forward scattering angles.
So measuring $P_\sigma$ for the $\gamma n$ reaction can give an
information on the size of the $K^* N \Theta$ interaction.
If it is close to $+1$, the $K^* N \Theta$ coupling would be large and
$K^*$ exchange is the dominant process in $\gamma n$ reaction.
If it is close to $-1$, however, it implies small value of $g_{K^*N
\Theta}^{}$.
Therefore, measurement of this quantity can test the quark model
predictions on $g_{K^*N \Theta}^{}/g_{KN\Theta}^{}$ that has very
different values depending on the parity of $\Theta^+$.

\section{Summary and Discussion}

The recently observed $\Theta^+(1540)$ is believed to be a member of the
baryon antidecuplet, but its spin and parity are not confirmed yet.
Careful analyses on various $\Theta^+$ production processes are thus
expected to give informations on the properties of the $\Theta^+(1540)$
and test various models for hadron structure.
As a continuation of our efforts to understand the $\Theta^+$ production
processes \cite{OKL03a,OKL03c}, we have investigated cross sections and
spin asymmetries of the $\gamma + N \to \overline{K}^* + \Theta^+$ reaction in
this paper as a complementary process of $\gamma + N \to \overline{K}
+ \Theta^+$.
In this calculation, the spin of the $\Theta^+(1540)$ is assumed to be
1/2 and the results are obtained for different assumptions for the parity
of the $\Theta^+$.
The obtained cross sections for $\gamma + N \to \overline{K}^* + \Theta^+$
are found to be comparable to those for $\gamma + N \to \overline{K}
+ \Theta^+$.
Our results also show that the cross sections for even-parity $\Theta^+$
is much larger than those for odd-parity $\Theta^+$ at least by an order of
magnitude.

More solid information for the parity of the observed $\Theta^+(1540)$ could
be obtained from the measurements of spin asymmetries.
Indeed, within the uncertainties due to the model-dependence of the
production mechanisms and the lack of information on several coupling
constants, we found that some target-recoil double asymmetries are
sensitive to the parity of the $\Theta^+$ at forward scattering angles.
But it would be very hard to be measured experimentally because of the
difficulties in identifying the nucleon polarization in $\Theta^+$ decay
distribution, which is needed to know the helicity of the $\Theta^+$.
Furthermore, the spin asymmetries depend not only on the parity
of $\Theta^+$ but also on the dynamics of the production mechanisms.
Thus it would be necessary to measure various spin asymmetries on the
proton and neutron targets in order to study the production mechanisms
and the parity of $\Theta^+$.
Among various spin asymmetries, the parity asymmetry $P_\sigma$ in the
$\gamma n$ reaction can give a robust information for the dominant
$t$-channel exchange and can be a clean signal for the $K^* N \Theta$
interaction.
Once we know the strength of the $K^* N \Theta$ interaction at least
qualitatively, we can then learn more about the production processes and the
parity of $\Theta^+$ by measuring other spin asymmetries such as single
asymmetries and beam-target double asymmetries.
When combined with the quark model predictions which lead to $P_\sigma
\sim +1$ ($-1$) for positive (negative) parity $\Theta^+$ in $\gamma n$
reaction, estimation of the $K^*N\Theta$ interaction may also give a clue
for the parity of $\Theta^+$.
Such experiments should be possible at current experimental facilities.

\acknowledgments

We are grateful to V.~Burkert, H.~Gao, A.~Hosaka, V.~Kubarovsky,
T.-S. H. Lee, T.~Nakano, B.-Y. Park and E.~Smith for fruitful discussions
and valuable informations.
Y.O. is grateful to the Thomas Jefferson National Accelerator Facility
for its warm hospitality.
This work was supported by the Brain Korea 21 project of Korean Ministry
of Education and by KOSEF under Grant No. 1999-2-111-005-5.
The work of Y.O. was supported in part by DOE contract DE-AC05-84ER40150
under which the Southeastern Universities Research Association (SURA)
operates the Thomas Jefferson National Accelerator Facility.

\end{document}